\documentclass[aps,pre,floatfix,twocolumn,groupedaddress]{revtex4-1}
\usepackage[dvips]{graphicx}
\usepackage{graphicx}
\usepackage{amsmath}
\usepackage{balance}
\usepackage{verbatim}
\usepackage[section]{placeins}
\usepackage{float}

\graphicspath{{Images/}}

\newcommand{\etalia}{{\it et al.}}
\newcommand{\la}{\left\langle}
\newcommand{\ra}{\right\rangle}

\newcommand{\br}{{\bf r}}

\newcommand{\PRL}{Phys.~Rev.~Lett.~}

\newcommand{\JCP}{J.~Chem.~Phys.~}

\newcommand{\JPCM}{J.~Phys.: Condens.~Matter~}

\begin{document}

\title{Influence of Solvent Quality on Conformations of Crowded Polymers}

\author{Wyatt J. Davis and Alan R. Denton}
\email[]{alan.denton@ndsu.edu}
\affiliation{Department of Physics, North Dakota State University, Fargo, ND 58108-6050, USA}

\begin{abstract}
The structure and function of polymers in confined environments, e.g., biopolymers in the cytoplasm 
of a cell, are strongly affected by macromolecular crowding. To explore the influence of solvent quality 
on conformations of crowded polymers, we model polymers as penetrable ellipsoids, whose shape fluctuations 
are governed by the statistics of self-avoiding walks, appropriate for a polymer in a good solvent. 
Within this coarse-grained model, we perform Monte Carlo simulations of mixtures of polymers and 
hard-nanosphere crowders, including trial changes in polymer size and shape. Penetration of polymers 
by crowders is incorporated via a free energy cost predicted by polymer field theory. To analyze the 
impact of crowding on polymer conformations in different solvents, we compute average polymer shape 
distributions, radius of gyration, volume, and asphericity over ranges of polymer-to-crowder size ratio 
and crowder volume fraction. The simulation results are accurately predicted by a free-volume theory 
of polymer crowding. Comparison of results for polymers in good and theta solvents indicates that 
excluded-volume interactions between polymer segments significantly affect crowding, especially in the 
limit of crowders much smaller than polymers. Our approach may help to motivate future experimental studies 
of polymers in crowded environments, with possible relevance for drug delivery and gene therapy.
\end{abstract}
\maketitle


\section{Introduction}\label{intro}
Polymer conformations can be influenced by macromolecular crowding~\cite{ellis2001a,ellis2001b}, 
which occurs when the volume accessible to a macromolecule is reduced by the presence of other 
macromolecules (crowders) in solution, or by geometric confinement imposed by hard 
boundaries~\cite{minton2001,ha2016}.
Over the past four decades, this phenomenon has drawn increasing interest within the biophysics community 
for its ubiquity in cellular and other biological environments~\cite{minton1980,minton1981,
minton2000,minton2005,richter2007,richter2008,elcock2010,hancock2012,denton-cmb2013}.
Inside cells, macromolecules occupy up to 20$\%$ of the volume of the cytoplasm and 
up to 40$\%$ of the volume of the nucleoplasm~\cite{vandermaarel2008,phillips2009}.  
Excluded-volume interactions with crowders influences the conformational and diffusional 
behavior of biopolymers (proteins, RNA, DNA) within cells~\cite{luo2017,ha2016} and 
can significantly modify biomolecular processes, such as protein folding~\cite{cheung2013}.  
Macromolecular crowding also has been implicated in promoting polymer aggregation 
associated with cataract formation~\cite{stradner2007} and in the pathogenesis of 
neurodegenerative diseases, such as Alzheimer's disease~\cite{mittal2015}.  
Closely related is the confinement of polymers by nanoparticles in nanocomposite 
materials~\cite{han2001,kramer2005a,kramer2005b,kramer2005c,balazs2006,mackay2006,richter2010}.

The impact of crowding on biopolymer structure and function has been investigated in a 
variety of modeling and experimental studies~\cite{goldenberg2003,dima2004,cheung2005,
wittung-stafshede2012,linhananta2012,denesyuk-thirumalai2011,denesyuk-thirumalai2013a,
denesyuk-thirumalai2013b,pincus-thirumalai2015,gorczyca2015a,gorczyca2015b,gagarskaia2017,yethiraj2017}.
For example, in a series of computational studies~\cite{denesyuk-thirumalai2011,
denesyuk-thirumalai2013a,denesyuk-thirumalai2013b,pincus-thirumalai2015}, the structure 
and function of RNA were shown to be strongly influenced by crowding.
Recent experiments combined fluorescence microscopy and particle tracking methods to show
that DNA conformations and mobility respond to crowding by dextran (mimicing cytoplasm 
conditions)~\cite{gorczyca2015a,gorczyca2015b}.  Other experiments showed that crowding by dextran 
or polyethylene glycol (PEG) influences the stability and folding of actin~\cite{gagarskaia2017}.
Diffusion NMR and neutron scattering were used to probe the size (radius of gyration) 
of PEG in aqueous solution with Ficoll 70 crowders~\cite{yethiraj2017}.
While many such studies have explored the impact of crowding on polymer size, 
relatively few thus far have addressed the influence of crowding on polymer shape.

The significance of aspherical polymer conformations was recognized by Kuhn~\cite{Kuhn1934}, 
who pointed out that the gross shape of a linear polymer, when viewed from a reference frame 
tied to the principal axes of the chain, matches that of an elongated, flattened ellipsoid.
A linear polymer chain can be realistically modeled by a random walk, whose steps are analogous 
to polymer segments.  Mathematical studies have demonstrated that an ensemble of random walks,
analogous to an ensemble of polymer conformations, can be characterized by statistical 
distributions of size and shape~\cite{fixman-stockmayer1970,solc1971,solc1973,theodorou1985,
rudnick-gaspari1986,rudnick-gaspari1987,bishop1988,bishop1991}.  The conformations of a 
random walk (or polymer chain) can be quantified by the gyration tensor, whose eigenvalues 
(in the principal axis frame) determine the radius of gyration and the principal radii 
of a general ellipsoid.

Early statistical mechanical studies of an ideal polymer (i.e., in a theta ($\theta$) solvent),
modeled as a freely-jointed, segmented chain or a random walk (RW) of independent steps, 
yielded accurate approximations~\cite{fixman1962,flory-fisk1966,fisher1966,flory1969}, and an exact 
expression~\cite{fujita1970,yamakawa1970}, for the probability distribution of the radius of gyration.
Subsequent simulation studies produced accurate fitting formulas for the distributions of the 
gyration tensor eigenvalues of ideal (RW) polymers~\cite{sciutto1994,murat-kremer1998,eurich-maass2001}
and nonideal, self-avoiding walk (SAW) polymers~\cite{sciutto1996}.  
 
The influence of solvent on excluded-volume interactions between segments of a polymer
in a dilute solution, and in turn on scaling with segment number $N$ (or molecular weight) of the 
average radius of gyration $R_g$ is well understood~\cite{deGennes1979,lhuilier1988}.
A polymer in a $\theta$ solvent, modeled by a random walk, has $R_g\sim N^{1/2}$, while a 
polymer in a good solvent, modeled by a self-avoiding walk, has $R_g\sim N^{3/5}$ (roughly).
For polymers in crowded environments, however, the influence of solvent on 
size and shape distributions is relatively poorly understood~\cite{vlassopoulos2015}.

In previous work, we studied the influence of nanoparticle crowding on the conformations of 
polymers in $\theta$ solvents -- specifically, the radius of gyration distribution within a 
spherical polymer model~\cite{lu-denton2011} and the shape distribution within the ellipsoidal 
polymer model~\cite{lim-denton-JCP2014,lim-denton-JCP2016,lim-denton-SM2016}.  The purpose of 
the present paper is to investigate the influence of solvent quality (good vs.~$\theta$) on the 
conformations of crowded polymers.  By comparing data from molecular simulations with predictions 
from free-volume theory, we demonstrate the importance of solvent quality for the sizes and shapes
of crowded polymers.  In Sec.~\ref{models}, we review the coarse-grained model of polymers
as soft, penetrable ellipsoids.  In Sec.~\ref{methods}, we outline our computational methods:
Monte Carlo simulation and free-volume theory.  In Sec.~\ref{results}, we present and interpret
results for shape distributions and average geometric properties of crowded polymers in good solvents.
Finally, in Sec.~\ref{conclusions}, we conclude and suggest possible extensions of our work.

\section{Models}\label{models}

\subsection{Coarse-Grained Model of Polymer Coil}\label{coarse-grained model}

\begin{figure}[h]
\begin{center}
\includegraphics[width=\columnwidth]{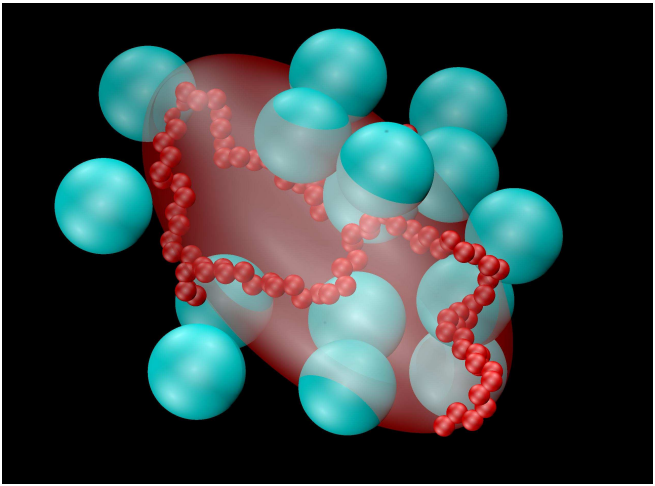}
\end{center}
\caption{Model of a linear polymer coil (red chain) approximated as a general ellipsoid 
that fluctuates in size and shape according to random-walk statistics and is 
penetrable by hard-sphere nanoparticles (blue spheres).
}\label{fig1}
\end{figure}

To efficiently explore the influence of nanoparticle crowding on the conformations 
of polymers in good solvents, we adopt a coarse-grained model of a polymer as a 
fluctuating ellipsoid whose shape distribution is governed by the gyration tensor 
of a self-avoiding walk (Fig.~\ref{fig1}):
\begin{equation}
{\bf T}~=~\frac{1}{N}\sum_{i=1}^N{\bf r}_i~{\bf r}_i~,
\label{gyration-tensor}
\end{equation}
where ${\bf r}_i$ is the position relative to the center of mass of segment (step) $i$
of $N$ total segments.
The eigenvalues of the gyration tensor -- $\Lambda_1$, $\Lambda_2$, $\Lambda_3$
in three dimensions -- determine the radius of gyration of the polymer in a particular conformation:
\begin{equation}
R_p=\left(\frac{1}{N}\sum_{i=1}^Nr_i^2\right)^{1/2}=\sqrt{\Lambda_1+\Lambda_2+\Lambda_3}~.
\label{Rp}
\end{equation}
(For reference, the gyration tensor relates to the moment of inertia tensor ${\bf I}$ 
via ${\bf T}=R_p^2{\bf 1}-{\bf I}$, with unit tensor ${\bf 1}$.)
The root-mean-square (rms) radius of gyration, which can be measured in scattering experiments, 
is given by
\begin{equation}
R_g=\sqrt{\la R_p^2\ra}=\sqrt{\la\Lambda_1+\Lambda_2+\Lambda_3\ra}~,
\label{Rg}
\end{equation}
where the angular brackets denote an ensemble average over polymer conformations.

If the ensemble average in Eq.~(\ref{Rg}) is defined relative to a frame of reference 
that rotates with the polymer's principal axes and, furthermore, the principal axes 
are labelled to preserve the order of the eigenvalues from largest to smallest 
($\Lambda_1>\Lambda_2>\Lambda_3$), then the average tensor describes an anisotropic 
object~\cite{rudnick-gaspari1986,rudnick-gaspari1987}.
The eigenvalues of the gyration tensor define an ellipsoid,
\begin{equation}
\frac{x^2}{\Lambda_1}+\frac{y^2}{\Lambda_2}+\frac{z^2}{\Lambda_3}=3~,
\label{ellipsoid}
\end{equation}
where $(x,y,z)$ are the coordinates of a point on the surface and the eigenvalues relate 
to the principal radii $R_i$ via $\Lambda_i=R_i^2/3$ ($i=1,2,3$).
This general ellipsoid is a coarse-grained representation of the average shape 
of the polymer (e.g., the tertiary structure of a biopolymer).
Each triplet of eigenvalues $\{\Lambda_1, \Lambda_2, \Lambda_3\}$
characterizes a unique polymer conformation and ellipsoid shape.  

In the absence of crowders, a three-dimensional SAW, modeling conformations of a 
linear polymer in a good solvent, has an average shape (eigenvalue) distribution 
determined by Monte Carlo simulations~\cite{sciutto1996} to be accurately 
described by a probability distribution,
\begin{equation}
P_0(\Lambda_1,\Lambda_2,\Lambda_3) = \prod_{i=1}^3 P_{i0}(\Lambda_i)~,
\label{P0}
\end{equation}
where the three factors are given by 
\begin{equation}
P_{i0}(\Lambda_i)=\frac{1}{\Gamma(\nu_i)} \frac{\nu_i}{\alpha_i}
\left(\frac{\nu_i\Lambda_i}{\alpha_i}\right)^{\nu_i-1}
\exp\left(-\frac{\nu_i\Lambda_i}{\alpha_i}\right)
\label{Pi0}
\end{equation}
and $\alpha_i$ and $\nu_i$ are fit parameters, tabulated in Table~\ref{table1}
for polymer chains of length $N=10^4$.  The factorized form of Eq.~(\ref{P0}) 
assumes independent eigenvalues -- aside from the ordering condition -- an assumption 
that proves accurate for sufficiently long polymers, with the exception of rare 
conformations in which an extreme extension in one direction can affect the 
probability of an extension in an orthogonal direction.  

\begin{table}
\caption{Parameters for shape distribution in Eq.~(\ref{Pi0})~\cite{sciutto1996}
for polymer chains of length $N=10^4$.}
\centering
\begin{tabular}{|c|c|c|c|}
\hline
eigenvalue $i$ & $\alpha_i$ & $\nu_i$ & $\Gamma(\nu_i)$ \\
\hline
\hline
1&7591.0120&3.35505&2.84226 \\
\hline
2&1604.3861&4.71698&15.8132 \\
\hline
3&544.16323&5.84822&92.8188 \\
\hline
\end{tabular}
\label{table1}
\end{table}

\begin{table}
\caption{Parameters for shape distribution in Eq.~(\ref{Pi0-scaled}).}
\centering
\begin{tabular}{|c|c|c|c|}
\hline
eigenvalue $i$ & $a_i$ & $b_i$ & $c_i$ \\
\hline
\hline
1& 11847.9 & 2.35505 & 22.3563 \\
\hline
2& 1.11669$\times 10^9$ & 3.71698 & 148.715 \\
\hline
3& 1.06899$\times 10^{14}$ & 4.84822 & 543.619 \\
\hline
\end{tabular}
\label{table2}
\end{table}

\begin{figure}[h]
\begin{center}
\includegraphics[width=\columnwidth]{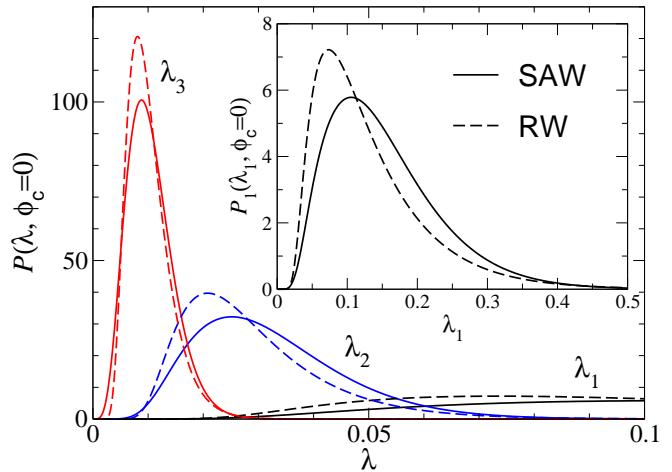}
\end{center}
\caption{Probability distributions $P(\lambda)$ of scaled gyration tensor eigenvalues 
$\lambda=\{\lambda_1,\lambda_2,\lambda_3\}$ of uncrowded polymers ($\phi_c=0$)
in a good solvent (SAW, solid curves) and in a $\theta$ solvent (RW, dashed curves).
Inset: largest eigenvalue $\lambda_1$ distributions.}
\label{fig2}
\end{figure}

An uncrowded SAW polymer of $N$ segments, each of (Kuhn) length $l$, has rms (average)
radius of gyration $R_g(0)=CN^{\nu}l$, with Flory exponent $\nu=0.588$ and amplitude 
$C=0.44108$~\cite{sciutto1996}.  For comparison, in a $\theta$ solvent, $R_g(0)=\sqrt{N/6}~l$.
Since the gyration tensor eigenvalues increase with $N$ in proportion to $N^{2\nu}$,
it is convenient to define scaled eigenvalues, $\lambda_i\equiv\Lambda_i/(N^{\nu}l)^2$,
in terms of which the shape distribution can be expressed as
\begin{equation}
P_{i0}(\lambda_i)=a_i \lambda_i^{b_i} \exp(-c_i\lambda_i)~,
\label{Pi0-scaled}
\end{equation}
where the parameters $a_i$, $b_i$, and $c_i$, derived from $\alpha_i$ and $\nu_i$,
are tabulated in Table~\ref{table2}.
The individual eigenvalue distributions differ somewhat from the factors in Eqs.~(\ref{Pi0})
and (\ref{Pi0-scaled}).  Each is obtained from the parent distribution [Eq.~(\ref{P0})] by 
integrating over the other two eigenvalues, with limits set by eigenvalue ordering
($\lambda_1>\lambda_2>\lambda_3$):
\begin{equation}
P_1(\lambda_1)=
\int_0^{\lambda_1}d\lambda_2\, \int_0^{\lambda_2}d\lambda_3\, P_0(\lambda)~,
\label{P1}
\end{equation}
\begin{equation}
P_2(\lambda_2)=
\int_{\lambda_2}^{\infty}d\lambda_1\, \int_0^{\lambda_2}d\lambda_3\, P_0(\lambda)~,
\label{P2}
\end{equation}
\begin{equation}
P_3(\lambda_3)=
\int_{\lambda_3}^{\infty}d\lambda_1\, \int_{\lambda_3}^{\lambda_1}d\lambda_2\, P_0(\lambda)~,
\label{P3}
\end{equation}
where $\lambda\equiv\{\lambda_1, \lambda_2, \lambda_3\}$ represents a triplet of scaled eigenvalues. 
For comparison, Fig.~\ref{fig2} shows the scaled eigenvalue distributions of uncrowded polymers
in good and $\theta$ solvents.
Note that, accounting for the different scaling factors -- $N$ for RW polymers, but $N^{1.176}$ 
for SAW polymers -- the unscaled eigenvalues are significantly larger for a SAW polymer in
a good solvent than for a RW polymer in a $\theta$ solvent.

Amidst crowders of volume fraction $\phi_c$, the rms radius of gyration $R_g(\phi_c)$ and 
the principal radii $R_i(\phi_c)$ are related to the scaled eigenvalues via
\begin{equation}
R_g(\lambda_1,\lambda_2,\lambda_3, \phi_c)=\frac{R_g(0)}{C}\sqrt{\la \lambda_1+\lambda_2+\lambda_3 \ra}
\label{Rg-over-Rg0}
\end{equation}
and
\begin{equation}
R_i(\lambda_i, \phi_c)=\frac{R_g(0)}{C}\sqrt{3\lambda_i}=3.9269R_g(0)\sqrt{\lambda_i}~.
\label{Ri}
\end{equation}
For comparison, in a $\theta$ solvent, $R_i=R_g(0)\sqrt{18\lambda_i}$ and
\begin{equation}
R_g(\phi_c)=R_g(0)\sqrt{6\la \lambda_1+\lambda_2+\lambda_3 \ra}.
\label{Ri-theta}
\end{equation}
The principal radii determine the average volume of the ellipsoidal polymer via 
\begin{equation}
v_p(\lambda_1,\lambda_2,\lambda_3, \phi_c)=\frac{4\pi}{3}\la R_1R_2R_3 \ra.
\label{vp}
\end{equation}

The deviation of a polymer's average shape from spherical is conveniently quantified 
by an asphericity parameter~\cite{rudnick-gaspari1986,rudnick-gaspari1987}, defined as
\begin{equation}
A=1-3\frac{\la\lambda_1\lambda_2+\lambda_1\lambda_3+\lambda_2\lambda_3\ra}
{\la(\lambda_1+\lambda_2+\lambda_3)^2\ra}~.
\label{asphericity}
\end{equation}
A perfect sphere has all eigenvalues equal and $A=0$, while an elongated object 
with one eigenvalue much larger than the others has $A\simeq 1$.
Crowding agents modify the eigenvalue probability distributions and, in turn,
the rms radius of gyration and asphericity of a polymer.

As in our previous studies of crowding of RW polymers in a $\theta$ solvent~\cite{lim-denton-JCP2014,
lim-denton-JCP2016,lim-denton-SM2016}, our model extends the classic Asakura-Oosawa-Vrij (AOV) 
model of colloid-polymer mixtures~\cite{asakura1954,vrij1976}, which idealizes nonadsorbing 
polymers as effective spheres of fixed size (radius of gyration).  Although qualitatively 
describing depletion-induced demixing of colloid-polymer mixtures, the AOV model completely 
neglects polymer conformational fluctuations, the influence of crowding on polymer size and shape,
and the penetrability of polymers by smaller colloids (nanoparticles).  We investigate
polymer crowding within an extended model of polymer-nanoparticle mixtures that includes
all of these features.

\subsection{Polymer Penetration Model}\label{polymer-crowding}

To allow for penetration of polymers by crowders, we further extend the AOV model, following 
previous work~\cite{lim-denton-JCP2014,lim-denton-JCP2016,lim-denton-SM2016,schmidt-fuchs2002},
by defining an average penetration free energy $\varepsilon$ that represents the average loss 
in conformational entropy of a polymer upon penetration by a hard-nanosphere crowder of radius $R_c$
(see Fig.~\ref{fig1}).
The dependence of $\varepsilon$ on $q$ can be motivated from a simple scaling argument
for an uncrowded polymer in the $N\to\infty$ ($q\gg 1$) limit~\cite{deGennes1979}.  
Given that $\varepsilon$ should be proportional to both the fraction of polymer volume 
occupied by the nanosphere and the number of polymer segments, we make the scaling ansatz,
\begin{equation}
\beta\varepsilon\sim \frac{R_c^3}{v_p}~Y(q)~,
\label{epsilon1}
\end{equation}
where the scaling function $Y(q)$ is proportional to $N$.  
Since $N\sim[R_g(0)]^{1/\nu}$ in a good solvent, it follows that $Y(q)\sim q^{1/\nu}$, which implies 
\begin{equation}
\beta\varepsilon\sim \frac{R_c^3}{v_p}~q^{1/\nu} \sim \frac{R_g^3(0)}{v_p}~q^{1/\nu-3}~,
\label{epsilon2}
\end{equation}
where $q\equiv R_g(0)/R_c$ is the uncrowded polymer-to-crowder size ratio.
For a SAW polymer in a good solvent at temperature $T$, polymer field theory~\cite{eisenriegler1996,
hanke1999,eisenriegler2000,eisenriegler2003} predicts $\nu=0.588$ and, in the limit
in which the crowders are much smaller than the polymers ($q\gg 1$):
\begin{equation}
\beta\varepsilon \simeq \frac{18.4 R_g^3(0)}{v_p q^{1.29932}}~.
\label{epsilon3}
\end{equation}
For $q\lesssim 1$, nanosphere insertions are so costly in free energy that the polymer is 
practically impenetrable.  Crowders then influence the polymer shape mostly from outside.  
Conversely, for $q\gg 1$, penetration is less costly and crowding can occur from both 
inside and outside the polymer.  If the polymer were approximated as a sphere of fixed radius, 
we would have $v_p=(4\pi/3)R_g^3(0)$ and
\begin{equation}
\beta\varepsilon \simeq \frac{4.4}{q^{1.29932}}~,
\label{epsilon4}
\end{equation}
an expression that has been used in previous studies of colloid-polymer mixtures in the 
$q\gg 1$ limit~\cite{bolhuis2003}.  In our study, however, we used the more general expression
of Eq.~(\ref{epsilon3}), which applies to a polymer of arbitrary shape.  In comparison, 
for a RW polymer, $\beta\varepsilon \simeq 4\pi R_g^3(0)/(qv_p)$, reducing to
$\beta\varepsilon \simeq 3/q$ in the spherical polymer model.

The coarse-grained ellipsoidal polymer model should be reasonable if, in the time required for
the crowders to significantly change their configuration, the polymer has sufficient time to 
equilibrate by visiting a representative sample of its possible conformations.  We must then 
assume a separation of time scales between diffusion of crowders and conformational 
rearrangement of the polymer.  For purposes of a rough estimate, we assume that a statistically 
independent configuration of crowders is achieved when each crowder diffuses a distance 
comparable to its diameter, while the polymer conformation equilibrates in the time it takes
for each segment to diffuse a distance comparable to the segment length.  Equating these distances 
-- presuming similar diffusion rates -- gives a lower limit on the allowed size of a crowder 
(upper limit on $q$) for which the model is reasonable.  This simple estimate yields 
$q\simeq\sqrt{N/6}$ for a RW polymer and $q\simeq CN^{\nu}$ for a SAW polymer.  Thus, for 
most practical purposes, and certainly for the systems considered in Sec.~\ref{results},
the coarse-grained ellipsoidal polymer model is quite justified.

\section{Methods}\label{methods}
\subsection{Monte Carlo Simulation}\label{MC}
Adapting methods developed in previous studies of polymer crowding~\cite{lim-denton-JCP2014,
lim-denton-JCP2016,lim-denton-SM2016}, we simulated a single ellipsoidal polymer, fluctuating 
in conformation according to the model described in Sec.~\ref{coarse-grained model}, immersed in a 
fluid of hard nanospheres.  In the canonical ensemble, with fixed numbers of particles 
at constant temperature in a cubic cell of fixed volume with periodic boundary conditions,
we implemented a variation of the Metropolis Monte Carlo (MC) algorithm.  Trial displacements of 
nanospheres and polymers were performed by translating the center of a particle at position 
$(x, y, z)$ to a new position $(x+\Delta x, y+\Delta y, z+\Delta z)$, where $\Delta x$, $\Delta y$, 
and $\Delta z$ were chosen independently and randomly in the range $[-0.2R_c, 0.2R_c]$.  

Trial displacements of polymers were coupled with trial rotations and shape changes as a single
composite move.  To ensure uniform sampling of the polymer orientation, defined by a unit vector 
${\bf u}$ aligned with the long axis of the ellipsoid, we generated a new orientation
${\bf u}_{\rm new}$ from an old orientation ${\bf u}_{\rm old}$ via~\cite{frenkel-smit2001} 
\begin{equation}
{\bf u}_{\rm new}=\frac{{\bf u}_{\rm old}+\tau{\bf v}}{|{\bf u}_{\rm old}+\tau{\bf v}|}~,
\label{rotation}
\end{equation}
where ${\bf v}$ is a randomly oriented unit vector and the tolerance $\tau$ was chosen randomly 
in the range $[-0.1, 0.1]$.  Trial variations in polymer shape were performed by changing one set of 
gyration tensor eigenvalues $\lambda_{\rm old}=\{\lambda_1, \lambda_2, \lambda_3\}$ to a new set 
$\lambda_{\rm new}=\{\lambda_1+\Delta\lambda_1, \lambda_2+\Delta\lambda_2, \lambda_3+\Delta\lambda_3\}$,
where $\Delta\lambda_1$, $\Delta\lambda_2$, and $\Delta\lambda_3$ were chosen independently and 
randomly in the ranges $[-0.01, 0.01]$, $[-0.003, 0.003]$, and $[-0.001, 0.001]$.  
A trial move (displacement, rotation, and shape change) was accepted with probability
\begin{equation}
{\cal P}_{\rm acc} = \min\left\{\frac{P_0(\lambda_{\rm new})}
{P_0(\lambda_{\rm old})} 
e^{-\beta\Delta F},~1\right\}~,
\label{shape-variation}
\end{equation}
where $\Delta F$ is the change in free energy resulting from the change in number of particle overlaps.
For the polymer-nanosphere penetration energy, we used Eq.~(\ref{epsilon3}) with the polymer volume
predicted by free-volume theory.  While in principle we should iterate until $v_p$ computed from the
simulation equals $v_p$ predicted by the theory, in practice the theory proved sufficiently
accurate that no iterations were required.
A trial move resulting in overlap of hard nanospheres (yielding infinite $\Delta F$) was immediately 
rejected.  A move that creates/eliminates a polymer-nanosphere overlap yields $\Delta F=\pm\varepsilon$
(see Sec.~\ref{polymer-crowding}).  

\begin{figure}[h]
\centering
\includegraphics[width=\columnwidth]{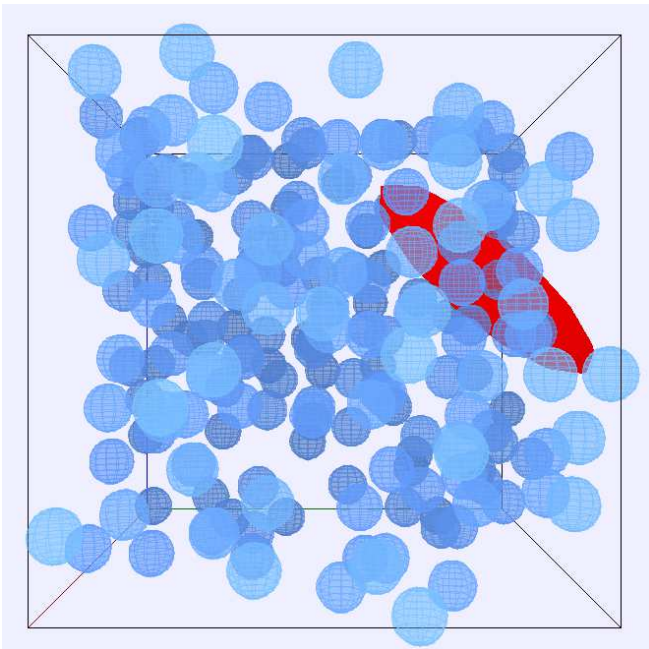}
\vspace*{-0.2cm}
\caption{
Snapshot of a simulation of $N_n=216$ nanospheres (blue spheres) and one polymer (red ellipsoid) 
in a cubic simulation cell.  The rms radius of gyration of the uncrowded polymer 
equals five times the nanosphere radius ($q$=5).}
\label{fig3}
\end{figure}

During a simulation, we checked whether a trial move resulted in overlap of a nanosphere 
with the polymer.  For each candidate -- identified as a nanosphere whose center lies 
inside a sphere centered on the ellipsoid of radius equal to the sum of the nanosphere 
radius and the longest principal radius of the ellipsoid -- we diagnosed overlap by 
computing the closest distance between the nanosphere center and the surface of the ellipsoid,
which involves computing the roots of a 6th-order polynomial~\cite{heckbert1994}.  
After an accepted change in polymer conformation, we reordered the eigenvalues by size. 
During each simulation, we averaged over MC steps, a step being defined as a trial displacement 
of every nanosphere and a trial change in polymer conformation, to compute ensemble averages of 
eigenvalue distributions and polymer geometric properties (see Sec.~\ref{results}).
We coded our simulations in Java within the Open Source Physics Library~\cite{osp-sip2006}.
Figure~\ref{fig3} shows a typical snapshot from a simulation. 

\subsection{Free-Volume Theory of Crowding}

To guide choices of system parameters and to help interpret our simulation results, we adapted a
free-volume theory previously developed for colloid-polymer mixtures~\cite{lim-denton-JCP2014,
lim-denton-JCP2016,lim-denton-SM2016}.  The theory generalizes the theory of Lekkerkerker 
\etalia~\cite{Lekk} from the AOV model~\cite{asakura1954,vrij1976} of hard, spherical polymers 
to a model of soft, penetrable, aspherical polymers, as described in Sec.~\ref{models}.

The free-volume theory approximates the mean value of a quantity $Q(\lambda)$, averaged over
polymer shapes, by
\begin{equation}
\la Q\ra=\int_0^{\infty}d\lambda_1\, \int_0^{\lambda_1}d\lambda_2\, \int_0^{\lambda_2}d\lambda_3\, 
Q(\lambda) P(\lambda, \phi_c)~,
\label{average}
\end{equation}
where
\begin{equation}
P(\lambda, \phi_c)=\frac{\alpha(\lambda, \phi_c)}{\alpha_{\rm eff}(\phi_c)} P_0(\lambda)
\label{P}
\end{equation}
is the shape probability distribution of the crowded polymer, $\alpha(\lambda, \phi_c)$ is the 
free-volume fraction of a polymer of shape $\lambda\equiv\{\lambda_1,\lambda_2,\lambda_3\}$ 
amidst hard nanospheres of average volume fraction $\phi_c$, and
\begin{equation}
\alpha_{\rm eff}(\phi_c)=\int_0^{\infty}d\lambda_1\, \int_0^{\lambda_1}d\lambda_2\, 
\int_0^{\lambda_2}d\lambda_3\, \alpha(\lambda, \phi_c) P_0(\lambda)
\label{alphaeff}
\end{equation}
is the effective free-volume fraction, averaged over polymer shapes.  Note that the limits 
of the eigenvalue integrals in Eqs.~(\ref{average}) and (\ref{alphaeff}) are chosen to 
respect the eigenvalue ordering ($\lambda_1>\lambda_2>\lambda_3$).  Previous studies of 
crowding of RW (ideal) polymers~\cite{lim-denton-JCP2014,lim-denton-JCP2016,lim-denton-SM2016}
did not impose eigenvalue ordering.  While this simple approximation proves very accurate and
efficient for RW polymers, it turns out to be less accurate for SAW polymers, increasingly so 
with increasing crowder volume fraction.

According to the Widom particle insertion theorem~\cite{widom1963}, the free-volume fraction 
is related to the average work $W$ required to insert a polymer of shape $\lambda$ into a 
sea of hard spheres of volume fraction $\phi_c$ via
\begin{equation}
\alpha(\lambda, \phi_c)=\exp[-\beta W(\lambda, \phi_c)]~.
\label{alpha1}
\end{equation}
The average insertion work or, equivalently, the free energy required to distort the 
hard-sphere fluid and create the volume and interfacial area sufficient to accommodate 
the polymer, can be expressed as
\begin{equation}
W(\lambda, \phi_c)=p(\phi_c)v_p(\lambda)+\oint_S dS\, \gamma(\phi_c, \br)~,
\label{W1}
\end{equation}
where $p$ and $\gamma$ are the pressure and interfacial tension of the hard-sphere fluid,
respectively, $v_p(\lambda)$ is the volume of the polymer, and the integral is over a closed surface $S$,
defined by the surface of the ellipsoidal polymer and parametrized by surface position vector $\br$.
This expression conveniently and conceptually separates thermodynamic properties of the 
hard-sphere fluid from geometric properties of the polymer.

Assuming a smooth interface between the polymer and the hard-sphere fluid, the interfacial tension,
which depends on the curvature of the interface -- dictated by the polymer shape --
can be expanded in powers of the mean curvature
\begin{equation}
K(\br)=\frac{1}{2}\left(\frac{1}{R_1(\br)}+\frac{1}{R_2(\br)}\right) 
\label{K}
\end{equation}
and the Gaussian curvature
\begin{equation}
H(\br)=\frac{1}{R_1(\br)R_2(\br)}~, 
\label{H}
\end{equation}
where $R_1(\br)$ and $R_2(\br)$ are the local radii of curvature at a point ${\bf r}$ on the interface.
Thus,
\begin{equation}
\gamma(\phi_c, \br)=\gamma_{\infty}(\phi_c)+\kappa(\phi_c)K(\br)+\bar\kappa(\phi_c)H(\br)+\cdots~,
\label{gamma}
\end{equation}
where $\gamma_{\infty}$ is the interfacial tension of a flat interface (with infinite radii of
curvature) and the coefficients $\kappa(\phi_c)$ and $\bar\kappa(\phi_c)$ are bending rigidities
of the hard-sphere fluid, which depend only on the hard-sphere volume fraction. 
Substituting Eq.~(\ref{gamma}) into Eq.~(\ref{W1}) and integrating over eigenvalues yields
the curvature expansion of the insertion work:
\begin{eqnarray}
W(\lambda, \phi_c)&=&p(\phi_c)v_p(\lambda)+\gamma_{\infty}(\phi_c)a_p(\lambda)
\nonumber\\[1ex]
&+&\kappa(\phi_c)c_p(\lambda)+2\pi\bar\kappa(\phi_c)+\cdots~,
\label{W2}
\end{eqnarray}
where $a_p(\lambda)$ is the surface area of the ellipsoidal polymer, 
\begin{equation}
c_p(\lambda)=\oint_S dS\, K(\br)
\label{cp}
\end{equation}
is the integrated mean curvature, and we have exploited the Gauss-Bonnet theorem: 
\begin{equation}
\oint_S dS\, H(\br)=2\pi\chi~,
\label{Gauss-Bonnet}
\end{equation}
where the Euler characteristic $\chi=1$ for an ellipsoid.
Substituting Eq.~(\ref{W2}) into Eq.~(\ref{alpha1}) and neglecting higher-order terms 
in the curvature expansion yields an approximation for the polymer free-volume fraction:
\begin{eqnarray}
\alpha(\lambda, \phi_c)&=&\exp\{-\beta[p(\phi_c)v_p(\lambda)+\gamma_{\infty}(\phi_c)a_p(\lambda)
\nonumber\\[1ex]
&+&\kappa(\phi_c)c_p(\lambda)+2\pi\bar\kappa(\phi_c)]\}~.
\label{alpha2}
\end{eqnarray}
In the limit of an infinitesimally small (point-like) polymer coil ($\lambda\to 0$), 
$v_p(\lambda)$, $a_p(\lambda)$, and $c_p(\lambda)$ all tend to zero and the free-volume fraction 
then reduces to $1-\phi_c$, implying that
\begin{equation}
\bar\kappa(\phi_c)=-\frac{k_{\rm B}T}{2\pi}\ln(1-\phi_c)~.
\label{limit}
\end{equation}
Finally, to incorporate the penetrability of the polymers, the crowder volume fraction is
replaced by an effective volume fraction, 
$\phi_c'=\phi_c(1-e^{-\beta\varepsilon})$~\cite{lim-denton-JCP2014,lim-denton-JCP2016,
lim-denton-SM2016,schmidt-fuchs2002}.
Combining Eqs.~(\ref{alpha2}) and (\ref{limit}) yields the final approximation
for the polymer free-volume fraction:
\begin{eqnarray}
\alpha(\lambda, \phi_c)&=&(1-\phi_c')\exp\{-\beta[p(\phi_c')v_p(\lambda)+
\gamma_{\infty}(\phi_c')a_p(\lambda)
\nonumber\\[1ex]
&+&\kappa(\phi_c')c_p(\lambda)]\}~,
\label{alpha3}
\end{eqnarray}
from which mean values follow via Eqs.~(\ref{average})-(\ref{alphaeff}).

With knowledge of $p(\phi_c)$, $\gamma_{\infty}(\phi_c)$, and $\kappa(\phi_c)$ in the
curvature expansion of the insertion work [Eq.~(\ref{W2})], the free-volume theory
can be implemented to predict the dependence of polymer size and shape on crowding
by a hard-sphere fluid.  For example, the radius of gyration $R_g(\phi_c)$ 
can be explicitly computed from
\begin{equation}
R_g(\phi_c)=\frac{R_g(0)}{C}\int d\lambda\,
P(\lambda, \phi_c) \sqrt{\lambda_1+\lambda_2+\lambda_3}~.
\label{Rg-phic}
\end{equation}
To compute the results reported in Sec.~\ref{results}, we used the 
accurate Carnahan-Starling expressions for the hard-sphere fluid 
properties~\cite{oversteegen-roth2005,hansen-mcdonald2006}:
\begin{eqnarray}
\beta p(\phi_c)&=&\frac{3\phi_c}{4\pi R_c^3}\frac{1+\phi_c+\phi_c^2-\phi_c^3}{(1-\phi_c)^3}
\nonumber \\[0.5ex]
\beta\gamma_{\infty}(\phi_c)&=&\frac{3}{4\pi R_c^2}\left[\frac{\phi_c(2-\phi_c)}{(1-\phi_c)^2}
+\ln(1-\phi_c)\right] \nonumber \\[0.5ex]
\beta\kappa(\phi_c)&=&\frac{3\phi_c}{R_c(1-\phi_c)}~.
\label{CSthermo}
\end{eqnarray}
For the integrated mean curvature, we numerically evaluated the integral in Eq.~(\ref{cp}) 
over a grid of eigenvalues and stored the results in a look-up table for later access.

\section{Results}\label{results}

\subsection{Simulation Protocol}\label{protocol}

\begin{figure}[t!]
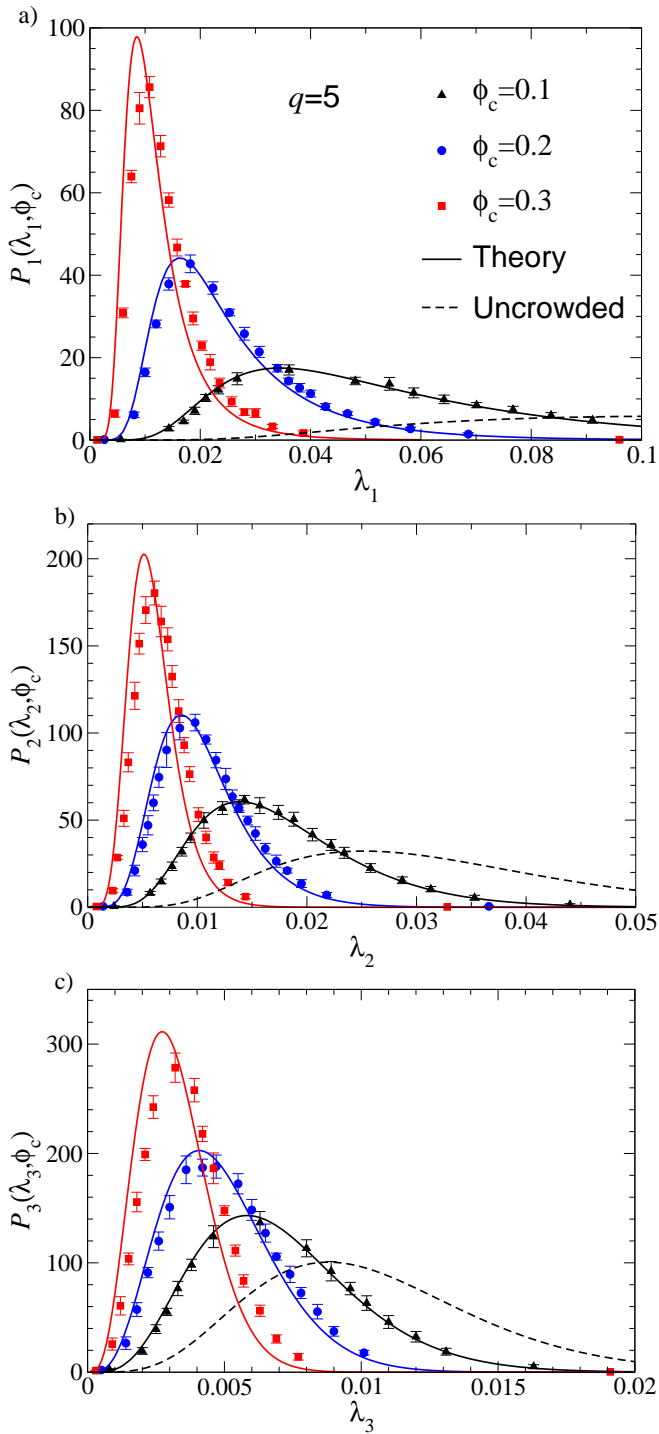

\includegraphics[width=\columnwidth]{L1_5_ordered_new.eps} 
\includegraphics[width=\columnwidth]{L2_5_ordered_new.eps}
\includegraphics[width=\columnwidth]{L3_5_ordered_new.eps}
\caption{Eigenvalue probability distributions ($\lambda_1>\lambda_2>\lambda_3$) of 
gyration tensor of a crowded polymer coil in a good solvent, modeled as a fluctuating, 
penetrable ellipsoid, governed by self-avoiding-walk statistics.
Simulation data (symbols) are compared with predictions of free-volume theory (solid curves) for a 
single polymer, with uncrowded rms radius of gyration equal to five times the nanoparticle radius 
($q=5$), amidst $N_n$ = 216 hard nanosphere crowders of volume fraction $\phi_c=0.1$ (triangles), 
0.2 (squares), and 0.3 (circles).  Dashed curves show uncrowded ($\phi_c=0$) distributions.}
\vspace*{5mm}
\label{fig4}
\end{figure}

\begin{figure}[t!]
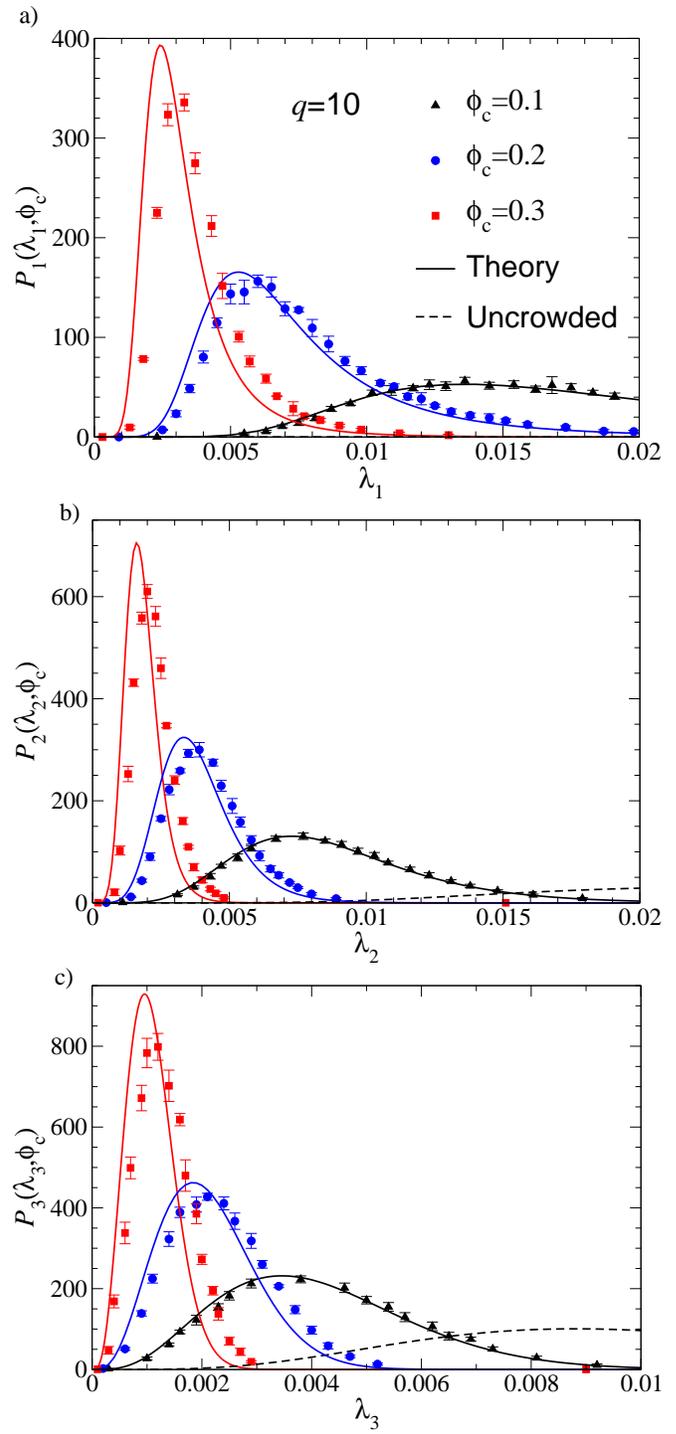

\includegraphics[width=\columnwidth]{L1_10_ordered_new.eps} 
\includegraphics[width=\columnwidth]{L2_10_ordered_new.eps}
\includegraphics[width=\columnwidth]{L3_10_ordered_new.eps} 
\caption{Same as Fig.~\ref{fig4}, but for uncrowded polymer-to-nanosphere size ratio $q=10$.
Notice changes in horizontal and vertical scales.}
\label{fig5}
\end{figure}

For several choices of uncrowded polymer-to-crowder size ratio $q$ and crowder volume fraction $\phi_c$, 
we performed simulations of a polymer coil (modeled as a fluctuating, penetrable ellipsoid) and $N_n$=216 
nanosphere crowders, initialized on a cubic lattice ($6\times 6\times 6$ array).  After an 
equilibration stage of 5 $\times$ $10^4$ MC steps, we collected statistics for the three eigenvalues 
of the gyration tensor at intervals of $10^3$ MC steps for $10^4$ intervals.  For each parameter 
combination, we ran five independent simulations and averaged over runs to obtain statistical errors.
We ran test simulations for longer times and larger systems to ensure that the system had reached
equilibrium and that finite-size effects were negligible.
From the raw eigenvalue data, we obtained probability distributions [Eq.~(\ref{P})] as histograms
(see Figs.~\ref{fig4} and \ref{fig5})
and computed average polymer geometric properties: radius of gyration [Eq.~(\ref{Rg-over-Rg0})], 
volume [Eq.~(\ref{vp})], and asphericity [Eq.~(\ref{asphericity})]
(see Figs.~\ref{fig6} and \ref{fig7}).

\clearpage

%

\clearpage

\subsection{Comparison of SAW and RW Polymers of Equal Uncrowded Size}\label{size-ratio}

We first present and discuss results that illustrate the influence of solvent quality on conformations
of crowded SAW and RW polymers having the same uncrowded size (radius of gyration).
Figures~\ref{fig4} and \ref{fig5} show our simulation results for the eigenvalue probability 
distributions for $q=5$ and $q=10$, respectively, over a range of crowder volume fraction.
Also shown are predictions of our free-volume theory.  Simulation and theory are in reasonable 
agreement, deviations increasing with $\phi_c$ as polymer-crowder correlations strengthen,
and both predict progressive shifts to lower eigenvalues (smaller principal radii) and narrowing
of the distributions with increasing crowder volume fraction.  These results reveal that, 
with increasing $\phi_c$, not only do polymer coils tend to contract along each principal axis,
but also fluctuations in size and shape are suppressed -- more so for $q=10$ than for $q=5$.  
The slightly larger deviations between theory and simulation at larger $q$ -- evident especially  
at higher $\phi_c$ -- reflect limitations of the theory associated with neglecting 
higher-order terms in the curvature expansion of the interfacial tension
[Eqs.~(\ref{gamma}) and (\ref{W2})].
These deviations propagate forward and affect the predicted size and shape of the crowded polymer, 
which depend on averages over the eigenvalue distributions.

\begin{figure}[t!]
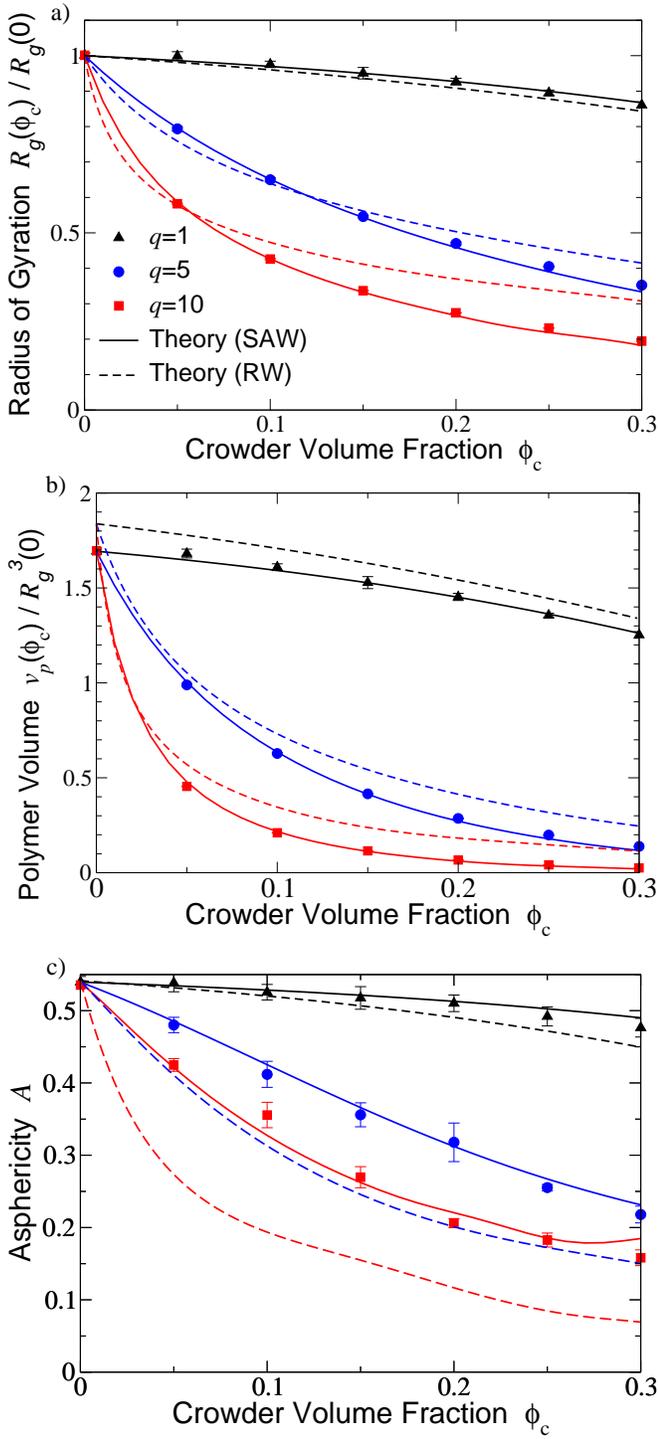

\includegraphics[width=\columnwidth]{Rg_ordered_new.eps} 
\includegraphics[width=\columnwidth]{Volume_ordered_new.eps}
\includegraphics[width=\columnwidth]{Asp_ordered_new.eps} 
\caption{Average geometric properties a polymer (modeled as a fluctuating, penetrable ellipsoid) 
vs. nanosphere crowder volume fraction $\phi_c$: (a) rms radius of gyration, (b) volume, (c) asphericity.
Simulation data are shown for SAW polymers of uncrowded polymer-to-nanosphere size ratio $q=1$ (triangles), 
$q=5$ (circles), and $q=10$ (squares).  For some points, error bars are smaller than symbols.
Corresponding predictions of free-volume theory are shown for SAW polymers (solid curves) 
and RW polymers (dashed curves).}
\label{fig6}
\end{figure}

Figure~\ref{fig6} illustrates the influence of crowding on geometric properties of a polymer coil.
The average radius of gyration, volume, and asphericity of the polymer all decrease monotonically 
with increasing crowder volume fraction.  Polymers of uncrowded $R_g$ equal to the crowder radius 
($q=1$) are relatively insensitive to crowding, experiencing only a 15\% reduction in size at 
$\phi_c=0.3$, while larger polymers ($q=5$, 10) contract significantly more than smaller polymers 
at the same $\phi_c$.  
The crowding effect predicted at $q=1$, though relatively weak, is stronger than that determined 
from the experiments of ref.~\cite{yethiraj2017}, in which no significant change in radius of gyration
was observed over the same range of $\phi_c$.  Although its source is unclear, this quantitative 
disagreement may result from some mismatch between our model and the experimental system.
It may be, for example, that the conformational statistics exhibited by the polymer (PEG) in water
are not quite those of a SAW, or that the interactions between crowding agents (Ficoll 70) differ 
from hard-sphere interactions, or that the polymer-crowder interactions are not purely entropic.

\begin{figure}[t!]
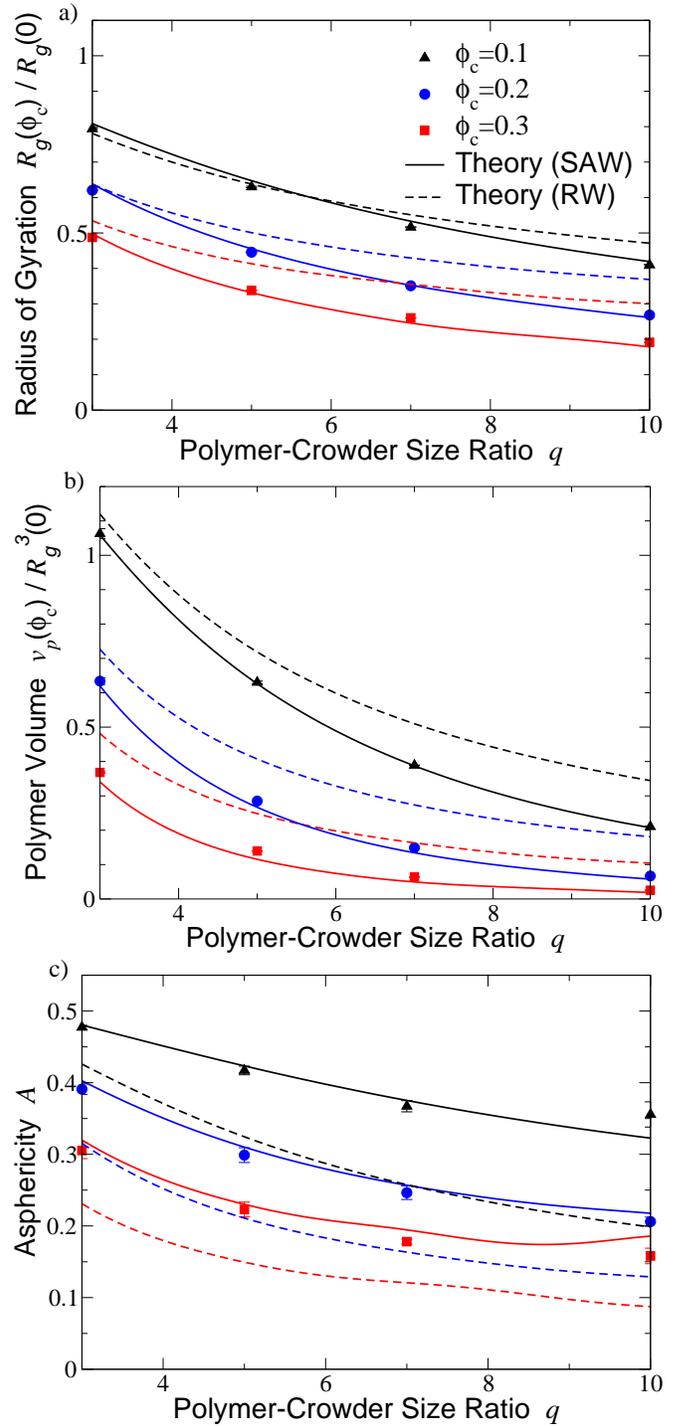

\includegraphics[width=\columnwidth]{RgvsQ_ordered_new.eps} 
\includegraphics[width=\columnwidth]{VolvsQ_ordered_new.eps}
\includegraphics[width=\columnwidth]{AspvsQ_ordered_new.eps} 
\caption{Average geometric properties of a polymer (modeled as a fluctuating, penetrable ellipsoid) 
vs. uncrowded polymer-to-nanosphere size ratio $q$: (a) radius of gyration, (b) volume, (c) asphericity.
Simulation data are shown for SAW polymers and nanosphere crowder volume fraction 
$\phi_c=0.1$ (triangles), $\phi_c=0.2$ (circles), and $\phi_c=0.3$ (squares).  
For some points, error bars are smaller than symbols.
Corresponding predictions of free-volume theory are shown for SAW polymers (solid curves) 
and RW polymers (dashed curves).}
\label{fig7}
\end{figure}

The tendency of polymer contraction to increase with increasing $q$, at fixed $\phi_c$, 
is a result of the free energy cost associated with penetration of a polymer by crowders.
Figure~\ref{fig7} illustrates more directly the variation of polymer geometry with size ratio, 
showing that radius of gyration, volume, and asphericity all decrease monotonically with 
increasing $q$.  Evidently, the larger the polymer relative to the crowder, the more severe 
the influence of crowding on polymer conformation.  This trend can be explained by the fact that 
the total penetration energy scales as $q^{1/\nu}=q^{1.7}$ for a given $\phi_c$.  The latter scaling 
follows from the scaling of the penetration energy as $q^{1/\nu-3}$ [Eq.~(\ref{epsilon2})] 
and the number of penetrating nanospheres per polymer as $q^3$.  While our results for the 
dependence of $R_g$ on $\phi_c$ and $q$ are qualitatively consistent with the measurements of
Palit~\etalia~\cite{yethiraj2017} for aqueous solutions of PEG and Ficoll 70, 
in that polymer contraction increases monotonically with increasing $q$ and $\phi_c$, 
quantitative differences exist.  Future studies may determine whether the discrepancies can be 
accounted for by the particular chain statistics of PEG in water or by non-hard-sphere interactions 
between the Ficoll 70 crowders or by other limitations of our model. 

Also shown in Figs.~\ref{fig6} and \ref{fig7} are corresponding predictions of free-volume theory 
for both SAW and RW polymers.  Results for SAW polymers from simulation and theory 
generally agree closely, as found also in previous studies of RW polymer~\cite{lim-denton-JCP2014,
lim-denton-JCP2016,lim-denton-SM2016}.  For clarity of presentation, we do not include here 
previously reported simulation data for RW polymers, as they are in near-exact agreement with simulation.
As noted in Sec.~\ref{MC}, the output values of $v_p$ are sufficiently close to the input values 
(from free-volume theory) that no iterations between simulation and theory were required.
We emphasize that imposing ordering of eigenvalues turns out to have negligible effect on the 
results for RW polymer, but is more significant for SAW polymer, especially at higher $\phi_c$.

%

While polymers in different solvents show qualitatively similar responses to crowding, all geometric 
measures decreasing with increasing $\phi_c$, there are significant quantitative differences.
For the same $q$ and $\phi_c$, a SAW polymer (in a good solvent) has a consistently smaller volume 
[Figs.~\ref{fig6}b and \ref{fig7}b] and higher asphericity [Figs.~\ref{fig6}c and \ref{fig7}c] 
than a RW polymer (in a $\theta$ solvent).  Thus, a SAW polymer in a crowded environment is 
more compressed and more elongated than a RW polymer of the {\it same uncrowded size} (same $q$).  
The comparison of crowded sizes of SAW and RW polymer is somewhat more complicated.  
Fig.~\ref{fig6}a shows that for $q=1$ the SAW polymer is consistently less contracted
over the whole range of crowder volume fraction, while for $q=5$ and $q=10$, the SAW polymer 
contracts less at lower $\phi_c$, but more at higher $\phi_c$.  This cross-over with increasing 
$q$ and $\phi_c$ in the degree of contraction of SAW and RW polymers reflects a complex
interplay between, on the one hand, chain statistics and conformational entropy, and on the
other hand, penetration free energy (see Sec.~\ref{polymer-crowding}).
When we consider, however, polymers of the {\it same segment number} (molecular weight), 
rather than the same uncrowded size, the responses of SAW and RW polymers to crowding 
are more distinct, as we discuss in the next section.

\subsection{Comparison of SAW and RW Polymers of Equal Segment Number}\label{segment-number}

Thus far, when comparing the crowded conformations (sizes and shapes) of a SAW polymer in a 
good solvent with those of a RW polymer in a $\theta$ solvent, we have considered polymers 
of the same uncrowded radius of gyration (same $q$ value).  Since the scaling of $R_g(0)$ 
with segment number $N$ depends on the solvent quality~\cite{deGennes1979} 
(see Sec.~\ref{coarse-grained model}), we have implicitly been comparing polymers of different 
segment numbers.  Our predictions, while of fundamental interest and qualitatively consistent 
with observed trends, may be difficult to compare with experiments in which 
polymers of the same molecular weight are studied under different solvent conditions.  
To facilitate more direct comparisons with experiments, we now compare predictions for 
polymers of equal segment number in good and $\theta$ solvents.

Given the segment (Kuhn) length $l$ of a linear polymer coil, the radius of a spherical crowder,
and the size ratio $q_{\scriptscriptstyle\rm RW}$ of a RW polymer in a $\theta$ solvent, 
the scaling relations (Sec.~\ref{coarse-grained model}) determine the size ratio 
$q_{\scriptscriptstyle\rm SAW}$ of a SAW polymer of equal segment number in a good solvent:
\begin{equation}
q_{\scriptscriptstyle\rm SAW}=C 6^{\nu}(R_c/l)^{2\nu-1}q_{\scriptscriptstyle\rm RW}^{2\nu}~.
\label{q-relation}
\end{equation}
As an example, motivated by the recent experiments of Palit \etalia~\cite{yethiraj2017}, we 
consider aqueous solutions of PEG and Ficoll 70 crowder.  Taking the Kuhn length of PEG in water 
-- considered good solvent conditions at room temperature -- as $l=7.6$ {\AA}~\cite{LEE20081590} 
(twice the persistence length), the radius of Ficoll 70 as $R_c=55$ {\AA}~\cite{vandenBerg3870}, 
and $q_{\scriptscriptstyle\rm RW}=3$,
we obtain $q_{\scriptscriptstyle\rm SAW}\simeq 7$ and $N\simeq 3000$ segments.
(For comparison, $q_{\scriptscriptstyle\rm SAW}=1$ translates into 
$q_{\scriptscriptstyle\rm RW}\simeq 0.6$ and $N\simeq 120$.)

\begin{figure}[h!]
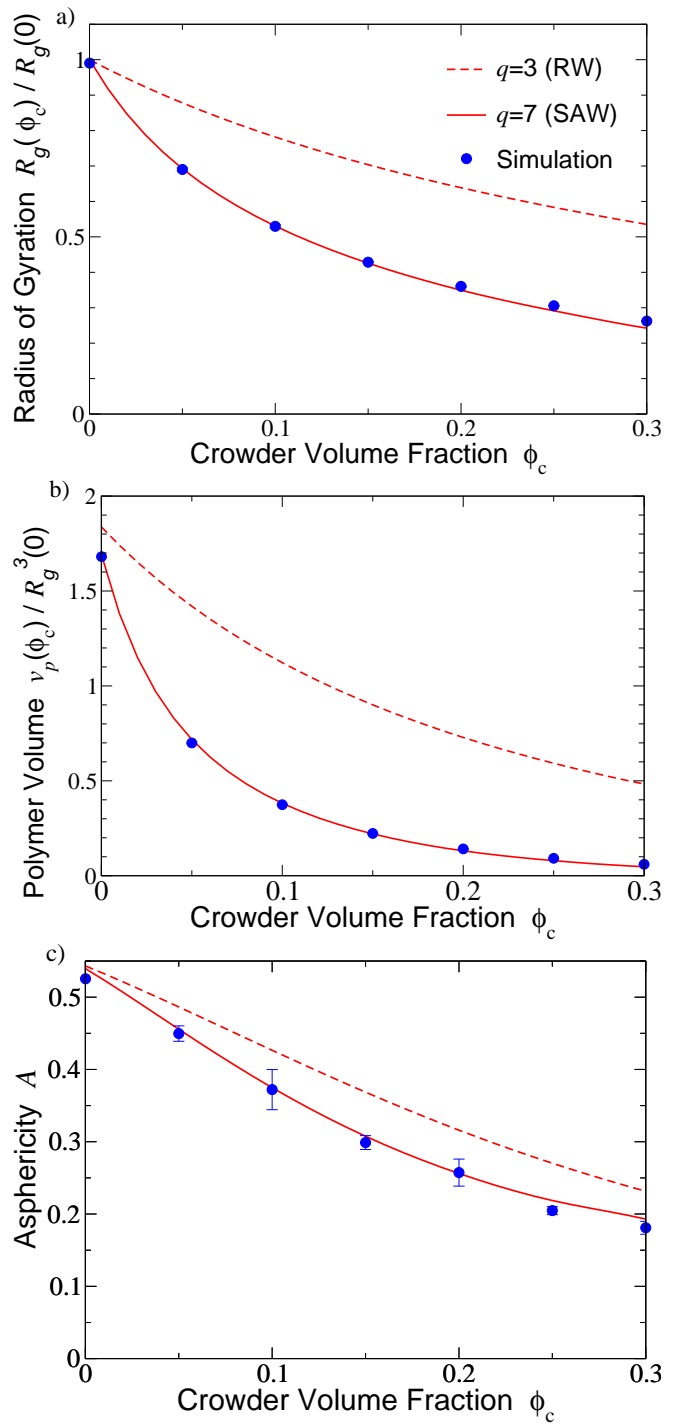

\includegraphics[width=\columnwidth]{PEGFicoll70_Rg_new.eps} 
\includegraphics[width=\columnwidth]{PEGFicoll70_Vol_new.eps}
\includegraphics[width=\columnwidth]{PEGFicoll70_Asp_new.eps} 
\caption{Average geometric properties of PEG (modeled as a fluctuating, penetrable ellipsoid)
vs. volume fraction $\phi_c$ of Ficoll 70 crowders (modeled as hard spheres) in water. 
(a) radius of gyration, (b) volume, and (c) asphericity of the polymer.
Simulation data (symbols) and theoretical predictions (solid curves) for SAW model 
of a polymer in a good solvent with $q\simeq 7$ are compared with predictions for
RW model of the same polymer in a $\theta$ solvent with $q=3$ (dashed curves). 
For both cases, the polymer contains $N\simeq 3000$ segments,
comparable to the experiments of ref.~\cite{yethiraj2017}.
}
\label{fig8}
\end{figure}

Figure~\ref{fig8} compares the geometric properties of this polymer in good and $\theta$ solvents. 
With increasing crowder volume fraction, the influence of crowding on 
conformations of a polymer of a given molecular weight --
quantified by radius of gyration, volume, and asphericity -- is consistently stronger in a good solvent 
than in a $\theta$ solvent.  These trends are not surprising, given our general understanding 
that crowding effects become more prominent with increasing $q$, and considering that 
$q_{\scriptscriptstyle\rm SAW}$ here exceeds $q_{\scriptscriptstyle\rm RW}$ by more than a factor of two.
What is perhaps most notable is that the average polymer shape (asphericity), while strongly 
dependent on crowder volume fraction, is relatively insensitive -- compared with average polymer size 
-- to a change in solvent quality.
It is worth noting that our $R_g$ vs. $\phi_c$ data have a sign of curvature (positive) that 
is opposite that of the experimental data~\cite{yethiraj2017}.  Quantitative comparisons are
complicated, however, by the relative simplicity of our model and the fact that, for given 
molecular weights of PEG, the measured uncrowded radii of gyration obey neither RW nor SAW 
chain statistics.

\section{Conclusions}\label{conclusions}
\vspace*{-3mm}
To summarize, we have investigated the dependence on solvent quality of polymer conformations 
in crowded environments.  For computational efficiency and conceptual simplicity, we modeled 
a polymer coil as an effective ellipsoid whose size and shape fluctuate according to the underlying 
statistics of random walks.  Specifically, fluctuations of the principal radii follow the 
probability distributions of the gyration tensor eigenvalues of either a self-avoiding walk --
for a polymer in a good solvent -- or a random walk -- for a polymer in a $\theta$ solvent.
Crowders are modeled as hard-sphere nanoparticles that mutually interact via a hard-sphere 
pair potential and are allowed to penetrate the volume enclosed by polymers with an average
free energy cost predicted by field theory.  

We implemented this coarse-grained model of polymer-crowder mixtures via Monte Carlo simulation 
and free-volume theory for polymers in good and $\theta$ solvents. 
As input to both simulations and theory, we used eigenvalue distributions previously determined 
from molecular simulations of random walks.  Our simulation data indicate that, with increasing 
crowder volume fraction and uncrowded polymer-to-crowder size ratio $q$, polymers become smaller 
and more compact (i.e., more spherical), in good agreement with predictions of free-volume theory.
While the conformations of SAW and RW polymers display similar qualitative trends, they exhibit 
significant quantitative differences.  For polymers that are equal in either radius of gyration 
or segment number, the influence of crowding is consistently stronger in a good solvent than in 
a $\theta$ solvent.  This dependence on solvent quality can be attributed to the important role
of segment self-avoidance on both the statistics and the penetrability of a polymer chain in 
a good solvent.  Our prediction can be experimentally tested by measuring the radius of gyration 
of polymers in dilute and crowded solutions under different solvent conditions, 
achievable by varying either temperature or concentration of a cosolvent.

The model considered here, in which polymer shapes are governed by random-walk statistics 
and crowders are treated as inert hard spheres, has the virtue of isolating and highlighting
the role of excluded volume in polymer crowding.  Quantitative description of many real systems,
however, may require incorporating intra-chain interactions (chain enthalpy), with appropriate 
uncrowded eigenvalue distributions, crowder-crowder interactions, and internal structure of crowders.
For example, real biopolymers may follow different chain statistics, while specific crowders
may be compressible and, if charged, may mutually interact by screened electrostatic potentials.  
Moreover, while our model of mobile polymers and crowders is designed to apply most closely
to macromolecular crowding of biopolymers in cellular environments, a model of polymers amidst 
fixed obstacles may be more applicable to polymer nanocomposite materials.  Future work should 
investigate the dependence of polymer conformations on chain statistics, crowder-crowder interactions, 
crowder structure, and crowder mobility.
Further simulations of more explicit (e.g., bead-spring) models of polymers in crowded 
environments~\cite{denesyuk-thirumalai2011,denesyuk-thirumalai2013a,denesyuk-thirumalai2013b,
pincus-thirumalai2015} also would help to test, calibrate, and refine the coarse-grained 
ellipsoidal polymer model.

\acknowledgments
This work was supported by the National Science Foundation under Grant No.~DMR-1106331.
Valuable discussions with Wei Kang Lim and Sylvio May and helpful correspondence with 
Sergio J.~Sciutto are gratefully acknowledged.




%

\end{document}